# Spontaneous emission into a planar optical waveguide mode by an atom outside the waveguide

Andrei Modoran and Gregory Lafyatis

The Department of Physics, The Ohio State University, Columbus, OH 43210

An electronically excited atom or molecule located outside but near a planar optical waveguide can decay by spontaneous emission of a photon into a guided mode of the waveguide. We outline a QED theory for calculating the probability for this process and describe general physical insights from that theory. A couple of representative examples are discussed in detail.

PACs numbers:42.50.Ct, 32.80.-t, 0.370+k

In this communication, we study the probability that an excited atom that is located just outside a planar optical waveguide will spontaneously decay by radiating a photon into a confined waveguide mode. *Our* interest in this problem stems from the suggestion by ourselves[1] and others[2,3] that ultra cold atomic samples can be trapped by strong evanescent wave optical frequency fields that may be created by driving planar waveguide modes. Planar waveguides are sometimes used as evanescent wave chemical and biological sensors[4] and our results may be relevant to those devices.[5,6] We also think the problem is fundamental enough to be intrinsically interesting.

Figure 1 defines the geometry and coordinates of the waveguides we consider. Urbach and Rikken[7] have formulated a basic QED treatment of planar waveguides and we will usually follow their conventions and notation. A planar optical waveguide is a thin film of a high index dielectric ($n_1$) sandwiched between two lower index regions. We consider an excited atom located in region 3, just above the waveguide's surface. For most of what follows, region 3 is vacuum, $n_3 = 1$. And $n_2$, is the index of the substrate of the waveguide film. We will also examine the case of a symmetric waveguide, for which $n_2 = n_3$.

An initially excited atom is located at position $\mathbf{R}$, just above the waveguide with the radiation field in its vacuum state: $|\Psi_i\rangle = |\psi_e, \mathbf{R}\rangle |0\rangle_R$. The "$R$" subscript denotes a state of the radiation field. Note that the atom's translational motion is here being treated classically. The rate for the atom to make a transition to a lower electronic state, $|\psi_g\rangle$ by spontaneously emitting a photon into a specific mode of the radiation field is given by Fermi's Golden Rule:

$$\frac{\delta w(E_f, \gamma_f)}{\delta \gamma} = \frac{2\pi}{\hbar} \left| v(E_f = E_i, \gamma_f; \Psi_i) \right|^2 \rho(E_f = E_i, \gamma_f) \qquad (1)$$

Here, $\gamma$ represents the parameters, e.g. wave vector coordinates and polarization, used in describing the radiation field mode. We discuss in more detail exactly what these parameters are, shortly. $\delta \gamma$ is a differentially small volume of parameter space centered on $\gamma_f$ and $\delta w(E_f, \gamma_f)$ is the rate for spontaneous emission for a photon into $\delta \gamma$. $E_i$ and $E_f$ are the initial and final energy of the atom + radiation field states. $\rho(E_f = E_i, \gamma_f)$ is the density of final states and $v(E_f = E_i, \gamma_f; \Psi_i)$ is the interaction Hamiltonian matrix element between the initial and final states of the system.

Explicitly, in the dipole approximation:

$$v(E_f = E_i, \gamma_f; \Psi_i) = \langle \psi_g |\,_R\langle (n=1:\gamma), 0... | -\hat{\mathbf{d}} \cdot \hat{\mathbf{E}}(\mathbf{R}) | \psi_e \rangle | 0... \rangle_R$$
$$= -\langle \psi_g | \hat{\mathbf{d}} | \psi_e \rangle \cdot \,_R\langle (n=1:\gamma_f), 0... | \hat{\mathbf{E}}(\mathbf{R}) | 0... \rangle_R \quad (2)$$

"hats" are used to identify operators. Here $\hat{\mathbf{d}} = -e \sum_i \hat{\mathbf{r}}_i$ is the electric dipole moment operator for the atom and its matrix element is the transition moment for the process. The electric field operator may be written in terms of a complete set of modes for the electromagnetic field of the system of Fig. 1:

$$\hat{\mathbf{E}}(\mathbf{R}, t) = \sum_\gamma i \left( \frac{\hbar \omega_\gamma}{2\varepsilon_0} \right)^{1/2} \mathbf{E}_\gamma(\mathbf{R}) e^{-i\omega_\gamma t} \hat{a}_\gamma + h.c. \quad (3)$$

$\mathbf{E}_\gamma(\mathbf{R})$ is a suitably normalized electric field function for the mode labeled by "$\gamma$," $\hbar \omega_\gamma$ is the photon energy for the mode, and the sum includes an integral over continuous parameters. The $\left( \frac{\hbar \omega_\gamma}{2\varepsilon_0} \right)^{1/2}$ factor arises from the normalization convention used by Urbach and Rikken.

In what follows we will study two limiting cases: (1) transitions with transition dipole matrix elements perpendicular to the plane of the waveguide: $\langle \psi_g | \hat{\mathbf{d}} | \psi_e \rangle$ has only a z-component and (2) transitions with transition matrix elements oriented parallel to the waveguide. As an example of the latter, consider a transition between two atomic states whose transition moment is oriented along the y-direction: $\langle \psi_g | \hat{\mathbf{d}} | \psi_e \rangle = -e y_{ge} \mathbf{j}$. "$\mathbf{j}$" is a unit vector in the y-direction and $y_{eg}$ is the matrix element of $\sum_i \hat{y}_i$, the y-displacements of the atomic electrons relative to the nucleus, taken between the two atomic states. The transition rate, Eq. 1 is proportional to the magnitude of the square of the interaction Hamiltonian matrix element of Eq. 2:

$$|v(E_f = E_i, \gamma_f; \Psi_i)|^2 = e^2 y_{eg}^2 \left(\frac{\hbar \omega_f}{2\varepsilon_0}\right) |E_{\gamma_f,y}(\mathbf{R})|^2 \qquad (4)$$

From Eq. 3, the only non-zero term in the mode expansion of the electric field matrix element of Eq. 2 is that corresponding to the mode receiving the emitted photon. And $E_{\gamma_f,y}(\mathbf{R})$ is the $y$ component of the normalized electric field wave function for that mode at the position of the atom. Thus *the rate for an atom to emit a photon into a confined waveguide mode (or any mode of the EM field) is proportional to the square of the electric field wave function of that mode at the atom's location.*

Next we outline those properties of optical waveguide modes that are relevant to this work and in particular describe the normalized electric field functions, $\mathbf{E}_\gamma(\mathbf{R})$, in Eq. 3. A complete set of modes with frequency, $\omega$ includes both free and guided wave solutions of Maxwell's equations. The guided waveguide modes are confined in the $z$-direction but may propagate within the film in any direction in the $x$-$y$ plane. The parameters, $\gamma$, describing a guided wave mode are, then, the propagation wavenumber components, $k_x$ and $k_y$, and the "Waveguide Mode" designation, $v$. "Waveguide Modes" are classified "TE" or "TM" depending on whether the mode's electric field or magnetic field is parallel to the waveguide's surface. For example, a TE mode that is propagating in the $x$-direction has its electric field in the $y$-direction. This holds both within the waveguide film and in the evanescent tails just outside the film where our atom is located. Similarly, a TM mode propagating in the $x$-direction has electric field components in both the $x$ and $z$ directions. The $x$ and $z$ components are out of phase by $90°$. The dependence of the electric and magnetic fields along the direction of propagation, is described by the mode's "propagation constant," $\beta = \sqrt{k_x^2 + k_y^2}$. E.g., for the TE1 mode traveling in the $x$-direction, $\mathbf{E}, \mathbf{B} \sim e^{-i\beta_{TE1}x}$. The propagation may also be characterized by the mode's "effective index;" e.g. $\beta_{TE1} \equiv n_{eff}^{TE1} k$, where $k = \frac{\omega}{c}$ is the wave vector of an electromagnetic wave of the same frequency propagating in vacuum. Assuming as in our case $n_2 \geq n_3$, the effective index for every mode is less than the index of the waveguide film, but greater than the substrate's, $n_1 > n_{eff} > n_2$. On both sides of the waveguide film the fields are evanescent. They fall off exponentially with distance from the surface. The characteristic decay constants for the evanescent fields are given by:

$$\kappa_j^\nu = k\sqrt{n_{eff,\nu}^2 - n_j^2} \tag{5}$$

Where $\nu$ is the waveguide mode designation and $j = 3$ or $2$ depending on whether one is above or below the waveguide. Thus, continuing the above, the electric field function appearing in Eq. 4, --- for a transition with a matrix transition moment in the $y$-direction that produces a photon in TE1 waveguide mode that propagates in the $x$-direction --- is:

$$E_{(TE1, k_x=\beta, k_y=0), y}(\mathbf{R}) = A e^{-i\beta X} e^{-\kappa_3^{TE1} Z} \tag{6}$$

Where $A$ is an overall normalization constant and $Y$ and $Z$ are the components of the atom's position. Modes are numbered such that lower order modes are more tightly confined to the film: $n_{eff}$'s and $\kappa$'s decrease with increasing waveguide orders.

To make a complete set of modes of the electromagnetic field for the infinite planar waveguide of Fig 1 the set of guided waveguide modes must be supplemented by modes that propagate freely in all directions. Urbach and Rikken describe one way to do this. The spontaneous emission rate into a particular waveguide mode, say TE1, is found by integrating Eq. 1 for that mode over all possible directions for propagation of the photon within the waveguide. Call this rate $w_{TE1}$. The total spontaneous emission rate of the excited atom, $w_{tot}$, is found by integrating Eq. 1 over *all* modes of the radiation field. The probability or branching ratio for emission into the particular waveguide mode is

$$P^{TE1} = \frac{w_{TE1}}{w_{tot}} \qquad (7)$$

As a representative first example we study the spontaneous emission of the Rb transition $5p_{3/2} \rightarrow 5s_{1/2}$ --- $\lambda = 780$ nm --- from an atom in vacuum just above a planar waveguide for the case of a transition moment parallel to the waveguide's surface. We consider a waveguide that is a 400 nm $Ta_2O_5$ film, $n_1 = 2.2$ on a fused silica substrate with, $n_2 = 1.45$. For $\lambda = 780$ nm this waveguide supports two TE modes and two TM modes. A transition with a dipole matrix element parallel to the surface can produce a spontaneous emission photon in any of these and Fig. 2 shows branching ratios into the individual modes. At the waveguide surface the probability for emitting into the waveguide is large --- approaching ½ in this case. Many features for the individual modes may be understood in terms of the normalized electric field wave functions for the guided modes, c.f. Eqs. (2), (4), and (6). The higher order modes are generally more extended into the region outside the waveguide film and thus have larger branching ratios. For the larger distances shown in Fig. 2 the asymptotically exponentially decreasing rates for emission into specific waveguide modes reflect the exponential fall off of modes in the evanescent wave above the surface: $w_\nu \sim e^{-2\kappa^\nu Z}$ away from the immediate surface. Indeed, the *rate* for spontaneous emission into a specific waveguide mode has this exponential dependence for *all* heights above the wave guide and the "flattening" of the *branching ratios*, very near the waveguide film's surface results from an *increase* in $w_{tot}$, Eq. (7).

"Enhanced (total) spontaneous emission" rate near a surface, has been the subject of several previous studies.[8] Finally note, these results are scalable in that if all of the lengths in the problem are multiplied by a common factor, corresponding branching ratios are identical for the two systems.

If we consider different waveguides with films having different thicknesses, as the thickness increases, the number of confined modes for a given frequency, $\omega$, also increases. We conclude by considering what happens near the birth of a waveguide mode. Modes are "born" on the substrate side of a waveguide in the sense that a waveguide whose thickness is just above that corresponding to the "appearance" of a new confined mode, will have for the new mode, $n_{eff} \approx n_2$, $\kappa_2 \to 0$ : the mode penetrates deeply into the substrate or, more generally, into that side of the film with the higher dielectric constant. This limiting behavior is reminiscent of the evanescent wave created by total internal reflection between two dielectrics near the critical angle. A symmetric waveguide, has long evanescent tails on *both* sides for a newly born mode.

We consider a Rb atom just above a symmetric, free-standing silicon nitride waveguide, $n_1 = 2.0$, $n_3 = n_2 = 1$. We examine emission of 780 nm radiation, this time from a transition whose dipole matrix element is oriented *perpendicularly* to the waveguide surface. A perpendicular transition emits only into the TM modes of the waveguide: recall, only those have a component of electric field in the z-direction. For symmetric waveguides the TM0 mode propagates for arbitrarily thin waveguides so we have chosen to study the birth of the TM1 mode. The minimum thickness waveguide that will support a TM1 mode is 225 nm. The magnitude of the evanescent wave is zero at the birth of a mode. Fig. 3 shows the branching ratio for spontaneous emission into the TM1 mode for three waveguides with thickness just above that of the TM1 mode's minimum. As expected, atoms much further from the waveguide than in the previous example emit with significant probability into the guided mode. Qualitatively, as the waveguide film increases in thickness, the emission into the mode grows in amplitude but extends less into the evanescent region. As before, very near the waveguide's surface, the overall "enhanced spontaneous emission" rate reduces the branching ratio into the TM1 mode. For this case, the extra emission is going mostly into the TM0 mode. To illustrate this, we have also shown in Fig. 3 the total branching ratio into both of the guided modes of the waveguide for a 255 nm film. At the waveguide's surface nearly 3/4ths the photons are emitted into confined modes.

In asymmetric waveguides, modes are born on the side of the film with the higher index of refraction. Thus, for a $Ta_2O_5$-fused silica waveguide of the first example, near the appearance of a new mode, *nothing* exceptional happens for an atom on the vacuum side of a waveguide. On the other hand, we consider a sensor consisting of a film on a low index substrate immersed in a higher index fluid. For waveguide thicknesses (or light wavelengths) near a mode's birth, collection by the waveguide of spontaneous emission photons on the fluid side will be substantially enhanced, especially for distances far into the fluid. For increasing thickness films, behavior qualitatively similar to that of the symmetric waveguide example is to be expected.

We acknowledge the support of the Research Corporation under grant No. RA0342.

# Figure Captions

**Fig. 1: Waveguide geometry and notation conventions. The excited atom is located in "region 3" just above the waveguide film.**

**Fig 2: Probability for spontaneous emission into guided modes of a waveguide as a function of the distance of an excited atom from the surface of the waveguide. Shown are branching ratios for spontaneous emission of a 780 nm photon into each allowed mode and the total probability for capture by the waveguide summed over the modes.**

**Fig. 3: Branching ratios for spontaneous emission into the TM1 mode for three waveguides with film thicknesses near 225 nm --- the minimum thickness that supports the mode. Also shown – the thin line--- is the total capture probability for a waveguide with thickness 255 nm.**

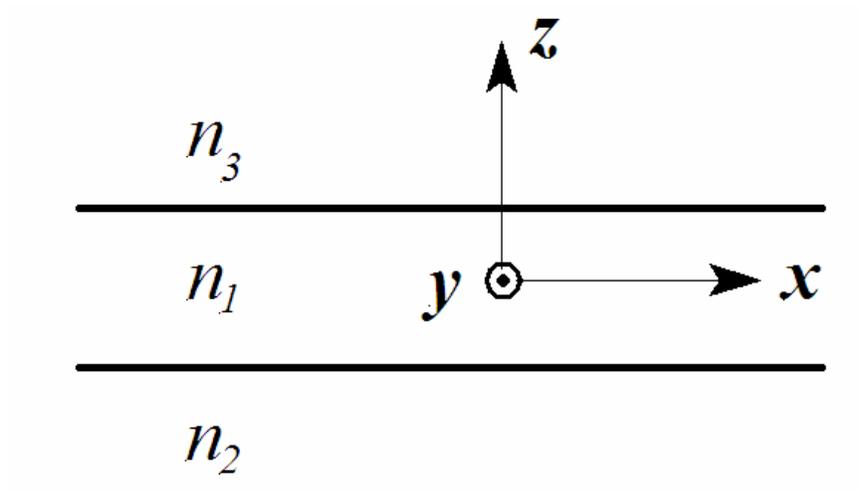

Figure 1

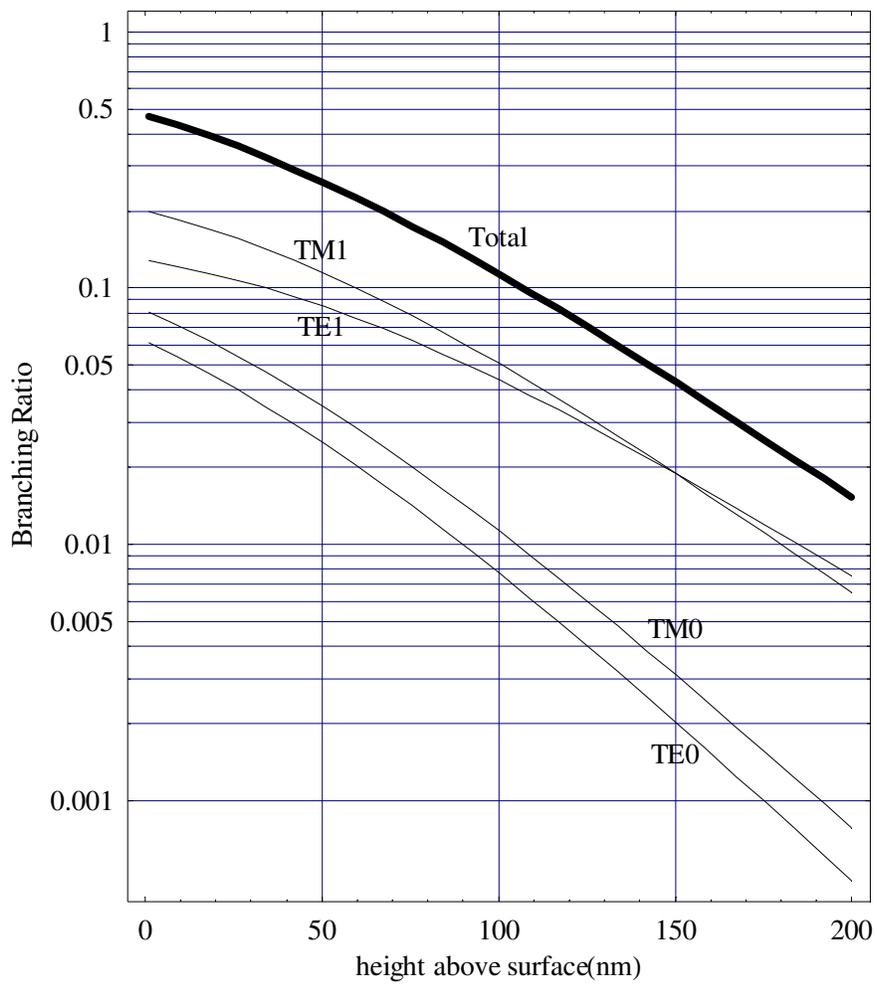

Figure 2

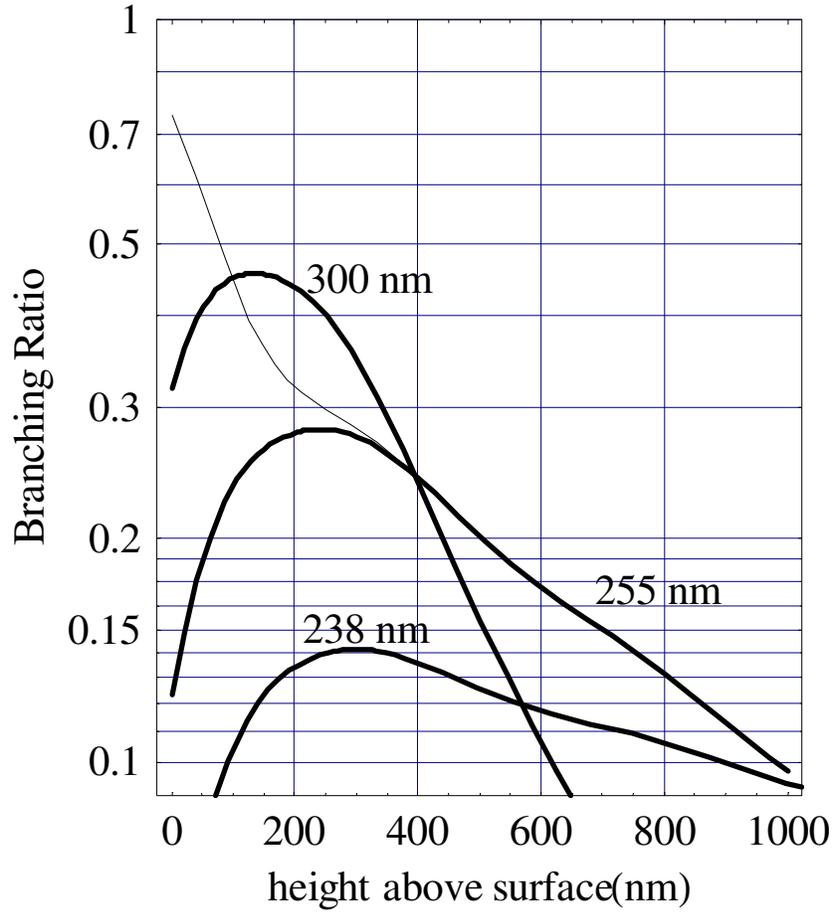

Figure 3